\documentclass[twocolumn]{aastex7}
\usepackage{nameref}
\usepackage{hyperref}
\usepackage{amsmath}
\usepackage{array}
\usepackage{enumitem}

\newcommand{\UCI}{Department of Physics \& Astronomy, The University of California, Irvine, Irvine, CA 92697, USA}

\begin{document}

\title{Hundreds of TESS exoplanets might be larger than we thought}

\author[0000-0002-7127-7643]{Te Han} 
\affiliation{\UCI}
\email[show]{tehanhunter@gmail.com}

\author[0000-0003-0149-9678]{Paul Robertson}
\affil{\UCI}
\email{paul.robertson@uci.edu}

\author[0000-0003-2630-8073]{Timothy D.~Brandt} \affiliation{Space Telescope Science Institute,
3700 San Martin Dr, Baltimore, MD 21218}
\email{tbrandt@stsci.edu}

\author[0000-0001-8401-4300]{Shubham Kanodia}
\affiliation{Earth and Planets Laboratory, Carnegie Institution for Science, 5241 Broad Branch Road, NW, Washington, DC 20015, USA}
\email{skanodia@carnegiescience.edu}

\author[0000-0003-4835-0619]{Caleb Ca\~nas}
\affil{NASA Postdoctoral Fellow}
\affil{NASA Goddard Space Flight Center, 8800 Greenbelt Road, Greenbelt, MD 20771, USA}
\email{c.canas@nasa.gov}

\author[0000-0002-1836-3120]{Avi Shporer}
\affil{Department of Physics and Kavli Institute for Astrophysics and Space Research, Massachusetts Institute of Technology, Cambridge, MA 02139, USA}
\email{shporer@mit.edu}

\author[0000-0003-2058-6662]{George Ricker}
\affil{Department of Physics and Kavli Institute for Astrophysics and Space Research, Massachusetts Institute of Technology, Cambridge, MA 02139, USA}
\email{grr@space.mit.edu}

\author[0000-0001-7708-2364]{Corey Beard}
\affil{\UCI}
\affil{NASA FINESST Fellow}
\email{ccbeard@uci.edu}

\begin{abstract}

The radius of a planet is a fundamental parameter that probes its composition and habitability. Precise radius measurements are typically derived from the fraction of starlight blocked when a planet transits its host star. The wide-field Transiting Exoplanet Survey Satellite (TESS) has discovered hundreds of new exoplanets, but its low angular resolution means that the light from a star hosting a transiting exoplanet can be blended with the light from background stars. If not fully corrected, this extra light can dilute the transit signal and result in a smaller measured planet radius. In a study of hundreds of TESS planet discoveries using deblended light curves from our validated methodology, we show that systematically incorrect planet radii are common in the literature: studies using various public TESS photometry pipelines have underestimated the planet radius by a weighted median of $6.1\% \pm 0.3\%$, leading to a $\sim20\%$ overestimation of planet density. The widespread presence of these biases in the literature has profoundly shaped—and potentially misrepresented—our understanding of the exoplanet population.  Addressing these biases will refine the exoplanet mass-radius relation, reshape our understanding of exoplanet atmospheric and bulk composition, and potentially inform prevailing planet formation theories.
\end{abstract}

\section{Introduction}
The Transiting Exoplanet Survey Satellite \citep[TESS;][]{Ricker2015} has discovered hundreds of exoplanets across the sky, measuring their radii from transit depths using its wide-field cameras. However, its low angular resolution introduces a bias that was less severe in its predecessor, the Kepler mission \citep{Borucki2010}: nearby stars often blend with exoplanet hosts, diluting the observed transits and leading to underestimated planet radii if not fully corrected. Inaccurate dilution corrections also lead to less reliable signal amplitudes for stellar variability and transients. 

The Robo-AO Kepler Survey revealed that $\sim 15\%$ of Kepler exoplanet hosts have unresolved companions within $4''$, which should have increased their planet radii by a factor of 1.08 on average if the primary is the host \citep{Law2014, Baranec2016, Ziegler2017, Ziegler2018, Ziegler2018b}. The Kepler Follow-up Observation Program concluded a $30\%$ occurrence rate of companions within the same range, causing an average correction factor of 1.06 assuming the primary star is the planet host \citep{Furlan2017}. For TESS, where blending is more pronounced due to lower angular resolution, the community has pursued follow-up observations of varying scales to quantify dilution effects \citep{Parviainen2020, Ziegler2020, Ziegler2021, Pelaez-Torres2024, Lillo-Box2024}. In particular, 117 of 542 TESS planet hosts observed with high-resolution imaging have close companions, which increases the radii of their planets by an average factor of 1.11 \citep{Ziegler2020}. These studies address statistical biases in a subset of systems with nearby companions, rather than systematic biases affecting the entire sample. However, the extended TESS point spread functions mean that background stars at several pixels away ($\sim 1'$) can still contaminate the exoplanet host. As a result, blending in TESS exists for almost all stars. Here, we present a statistical approach to evaluate the systematic impact of blending on TESS-derived planet radii using deblended light curves. 

The TESS-Gaia Light Curves \citep[TGLC;][]{Han2023} algorithm was developed to address the dilution problem in TESS. TGLC models the effective point spread function (PSF) of all stars in TESS full-frame images using star positions and magnitudes derived from Gaia Data Release 3 \citep{Gaia2023} as priors. It then forward models the field stars and subtracts their flux contributions from the target star. The resulting light curves exhibit less blending and are expected to accurately reflect signal amplitudes, such as those of planet transits. Other TESS light curves have applied different methods to tackle the blending problem. For example, the Science Processing Operations Center  pipeline \citep[SPOC;][]{Jenkins2016} uses a simulated, analytical pixel response function (PRF) to estimate the contamination ratio of a light curve. However, when applied to real TESS data, the PRF model—unlike the empirical effective PSF model used by TGLC—often fails to account for the irregularities of the shape of the TESS PSF, sometimes leading to inaccurate dilution corrections \citep{Lund2021,Parviainen2021}.

In a study testing the accuracy of TGLC signal amplitudes using TESS and Kepler exoplanets, we confirmed that TGLC has an accurate transit depth recovery. In addition, we identified a systematic underestimation of exoplanet radii in the literature for 228 TESS exoplanets by a weighted median of $\sim 6\%$. This widespread bias directly impacts the inferred compositions of TESS exoplanets, systematically skewing their densities by $\sim 20\%$ and distorting empirical mass-radius relations. Furthermore, it may also influence broader exoplanet demographics, atmospheric characterization priorities, and the apparent location of the evaporation valley.

This letter is structured as follows: In Section~\ref{sec:fit}, we recalculate the radii of hundreds of TESS exoplanets using TGLC light curves. Section~\ref{sec:bias} validates our methodology against Kepler exoplanets and quantifies the systematic underestimation of radii in the literature. Section~\ref{sec:discussion} explores the implications of these biases, including shifts in mass-radius relations, compositional interpretations, and population-level trends. We summarize our findings and their consequences for exoplanet studies in Section~\ref{sec:summary}.

\section{Reassessing TESS exoplanet radii with TGLC} \label{sec:fit}
To test the accuracy of the dilution correction of TGLC, we need to compare the exoplanet radii derived from TGLC to literature values. We first selected 346 confirmed exoplanets from the NASA Exoplanet Archive \citep[NEA;][]{Akeson2013} Planetary Systems table \citep{NEA_PST} for this analysis (retrieved on December 7, 2024). We selected exoplanets in single-planet systems with non-grazing transits—defined as those with an impact parameter plus planet-to-star radius ratio less than 1—to enable more reliable light curve modeling. This selection also ensures high signal-to-noise ratios, as 87\% of the sample are TESS Objects of Interest \citep[TOI;][]{Guerrero2021}, which typically require a signal-to-pink-noise ratio \citep[as defined by][]{Hartman2016} higher than 9. For planets with multiple measurements, we adopted the publication that had higher measured precision on the planet radii. Throughout our analysis, we use stellar and planetary parameters from a single publication per system to ensure consistency. In total, we collected 1723 TGLC single-sector light curves for 346 planets observed in TESS Sectors 1–55 \citep{tglc_hlsp}, meaning that most planets have light curves from multiple sectors.

We then ran TGLC-photometry-only light curve fits using the \textsf{exoplanet} package \citep{Foreman-Mackey2021} for each TESS sector with wide priors around the literature ephemeris and transit depth (Table~\ref{tab:prior}). This sector-by-sector fitting strategy was chosen deliberately to ensure a uniform and automated approach across the full sample of exoplanets, regardless of how many sectors each system appears in. While individual systems in the literature are often treated differently depending on available data or science goals, our aim is not to reproduce those results, but to provide an alternative measurement of the planet-to-star radius ratio ($p \equiv R_{\text{p}}/R_*$) to compare with literature values. Furthermore, fitting each sector independently serves an additional purpose: it allows us to test the stability of the radius ratio measurement under varying contamination conditions. As the TESS PSF, field orientation, and centroid position change from sector to sector, the contamination level varies accordingly. Thus, each sector offers an independent probe of photometric fidelity, particularly relevant for assessing the effectiveness of TGLC in mitigating contamination.

In modeling these transits, we also considered the degeneracies between planetary orbital eccentricity and impact parameter that arise in photometry-only fits. Moreover, the impact parameter directly influences the transit depth, and thus the inferred radius ratio $p$, on a limb-darkened stellar disk. To avoid biasing our measurements of $p$, we adopted the following strategy: for the 60 planets in our sample with eccentricities measured at $\geq$3$\sigma$ significance in the literature, we performed eccentric fits using Gaussian priors on both eccentricity and argument of periastron, centered on the literature values (Table~\ref{tab:prior}). For the remaining systems, whose orbits are consistent with circular, we adopted circular fits.

We also applied rotation-based Gaussian Process (GP) regression to each fit to remove long baseline trends. We used the $\texttt{RotationTerm}$ kernel from $\texttt{celerite2}$ \citep{Foreman-Mackey2017, Foreman-Mackey2018} to model the stellar variability, which consists of two damped simple harmonic oscillators. 800 of these sector-by-sector fits (of 271 planets) converged successfully with the $\hat{R}$ convergence diagnostic smaller than 1.01 for $p$. The rest were discarded to avoid manual tuning in order to preserve a homogeneous approach to all targets. The exclusion of these fits does not invalidate other successful fits. We also visually inspected each fit to ensure the correct transits are fitted. In particular, $\sim 20\%$ of the planets have orbital periods longer than 14 days, causing their transits to be absent in some TESS sectors and consequently leading to fitting failures. The fit results are hosted on Zenodo \citep{Han2025}.

\section{Systematic radius underestimation across hundreds of TESS exoplanets} \label{sec:bias}

\begin{figure}
    \centering
    \includegraphics[width=\columnwidth]{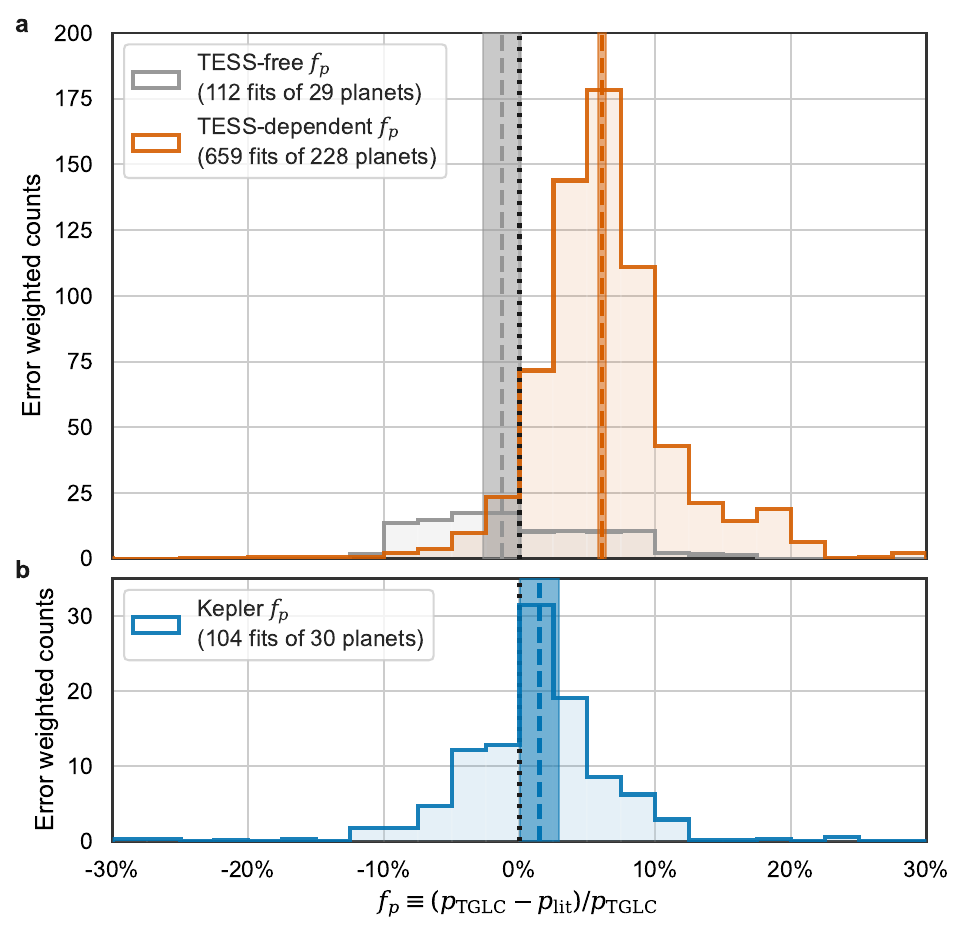}
    \caption{TESS-free, TESS-dependent, and Kepler literature planet-to-star radius ratios compared to fitted TGLC-fitted values. a, The inverse-variance weighted distribution of the fractional difference in radius ratios ($f_p$) between TGLC and literature values for TESS-free (gray) and TESS-dependent (orange) planets. b, The same distribution for Kepler planets (blue). The vertical dashed line and shaded region represent the medians of each distribution and the $1\sigma$ uncertainties of the medians estimated via bootstrap resampling. The black vertical dotted line is at $f_p = 0$ for reference. The TGLC-fitted radius ratios statistically agree with literature values for the less diluted TESS-free and Kepler planets, but disagree for TESS-dependent planets, implying a systematic underestimation of their radii in the literature.}
    \label{fig:radius_bias}
\end{figure}

To assess whether the recalculated TGLC exoplanet radii are consistent with literature values, we compare the planet-to-star radius ratios from TGLC and the literature. Specifically, we analyze the fractional difference
\begin{equation} \label{eq:1}
f_p \equiv (p_{\text{TGLC}} - p_{\text{lit}}) / p_{\text{TGLC}},
\end{equation}
where $p_{\text{TGLC}}$ is the radius ratio measured from TGLC light curves, and $p_{\text{lit}}$ is the value reported in the literature. This metric, $f_p$, reflects the fractional change in inferred planet radius and serves as an effective proxy for contamination. By normalizing the difference by $p_{\text{TGLC}}$, $f_p$ isolates the effect of contamination from differences in exoplanet transit depth, enabling a more direct comparison across systems. For example, two systems with different intrinsic transit depths but the same level of contamination will exhibit the same $f_p$, making it a suitable diagnostic for identifying potential dilution effects in the light curves. 

In order to calculate $f_p$ and propagate uncertainties, we require published errors on $p_{\text{lit}}$, or on both $R_{\text{p}}$ and $R_*$. However, 14 planets in our converged sample of 271 planets lack published errors. To maintain a homogeneous comparison, we therefore restrict our analysis to the 771 sector-by-sector fits of 257 planets.

\subsection{Radius ratio recovery accuracy differs between TESS-free and TESS-dependent planets}
Our analysis reveals a peak near 6\% for the distribution of the inverse-variance weighted $f_p$. The estimated $f_p$ for each planet-sector combination is inversely weighted by its variance propagated from $p_{\text{TGLC}}$ and $p_{\text{lit}}$, ensuring that more precise measurements contribute more significantly. The discrepancy between $p_{\text{TGLC}}$ and $p_{\text{lit}}$ led us to suspect that undercorrected contamination in the literature photometry may be the cause. To explore this, we grouped the TESS exoplanets based on the methodology used to derive their radii in the literature. In the ``TESS-free'' category (Table~\ref{tab:tess-free}), dilution in TESS photometry was treated as a free parameter, with radius ratios constrained solely by ground-based photometry. In the ``TESS-dependent'' category (Table~\ref{tab:tess-dependent}), radii were derived using TESS photometry, either alone or in combination with ground-based data. We separated and analyzed our sample according to these categories (Appendix~\ref{app:categorization}).  Using a total of 771 $p$ measurements from sector-to-sector fits, we computed the median $f_p$ of each category and estimated its uncertainty via bootstrap resampling, which randomly selects ten thousand new samples of the same size from the full dataset while allowing repetitive selection. The standard error is then calculated by taking the standard deviation of the weighted means of these samples. 

The resulting inverse-variance weighted median of $f_p$ is
\begin{equation}
f_{p,\,\mathrm{TESS\text{-}free}} = -1.3\% \pm 1.4\%
\end{equation}
for TESS-free planets ($p_{\text{lit}} = p_{\text{TESS-free}}$), and
\begin{equation}
f_{p,\,\mathrm{TESS\text{-}dependent}} = 6.1\% \pm 0.3\%
\end{equation}
for the TESS-dependent planets ($p_{\text{lit}} = p_{\text{TESS-dependent}}$). These results indicate that $p_{\text{TGLC}}$ statistically disagrees with $p_{\text{TESS-dependent}}$, but agrees with $p_{\text{TESS-free}}$ (Figure~\ref{fig:radius_bias}), which hints at an accurate signal amplitude recovery with TGLC. 

We also tested whether the discrepancy could be explained by grouping the TESS sample by other parameters, including stellar magnitude (which strongly correlates with photometric precision), orbital period, and light curve cadence. We divided the 771 fits into equal-sized subsets by magnitude (brighter vs. fainter), orbital period (longer vs. shorter), and cadence (1800s vs. 600s). The resulting three pairs of distributions and their medians were statistically indistinguishable, with differences below 1$\sigma$ significance. These tests indirectly support the interpretation that the TESS-free versus TESS-dependent nature of the radius measurements is the likely cause of their differences in $f_p$.

\subsection{Validating TGLC Signal Amplitude Accuracy Using Kepler Planets}
While the agreement with TESS-free planets implies unbiased radius measurements from TGLC, the ground-based photometries required for TESS-free radius measurements have limited precision. To provide additional evidence that TGLC has an accurate signal amplitude measurement, we added a similar test comparing TGLC-fitted radii to Kepler \citep{Borucki2010} planet radii, using TESS observations of exoplanets originally discovered by Kepler. 

In this test, we used TESS data processed with TGLC for exoplanets that were originally discovered and characterized by Kepler, not Kepler data itself. We queried the NEA Planetary Systems table on February 18, 2025 and limited the search to planets discovered by Kepler. We limited the sample to single planet systems with non-grazing transits and required the period to be less than 14 days for TESS sectors to effectively capture the transits. We only selected transits of 0.5\% or deeper and planet hosts with TESS magnitude 16 or brighter for TESS photometry to possibly resolve the transit \citep{Han2023}. A total of 53 Kepler planets from the NEA are selected by the above criteria. We collected a total of 281 light curves from TESS Sectors 1-55. Following the same methods and result selections in Section~\ref{sec:fit}, we successfully fit 104 light curves of 30 planets (Table~\ref{tab:kepler}). We adopted circular orbits for all Kepler planets, since they are all reported to have eccentricities measured at $<$3$\sigma$ significance. The inverse-variance weighted median of $f_p$ for the Kepler sample is consistent with zero within about one standard deviation (Figure~\ref{fig:radius_bias}), with a measured value of
\begin{equation}
f_{p,\,\mathrm{Kepler}} = 1.47\% \pm 1.43\%.
\end{equation}
Notably, accounting for the median $\sim1\%$ radius underestimation in the Kepler sample \citep{Furlan2017} brings $p_{\text{TGLC}}$ into even closer agreement with the expected true radius ratios, lending further confidence to its accuracy.

In contrast, the TGLC-fitted planet radii show a statistically significant discrepancy when compared to literature planet radii of the TESS-dependent planets. Given that TGLC applies a consistent methodology to all light curves, this discrepancy further implies a systematic bias for the TESS-dependent radius measurements. A positive $f_p$ for TESS-dependent planets suggests insufficient contamination removal in light curves reported in the literature for these systems. We conclude that the literature values of 228 TESS-dependent planets underestimate $p$ and thus planet radii by $\sim 6\%$. 

\section{Implications for exoplanet studies} \label{sec:discussion}
\subsection{Revised TESS exoplanet mass-radius-density distributions and compositional shifts}
\begin{figure*}
    \centering
    \includegraphics[width=\textwidth]{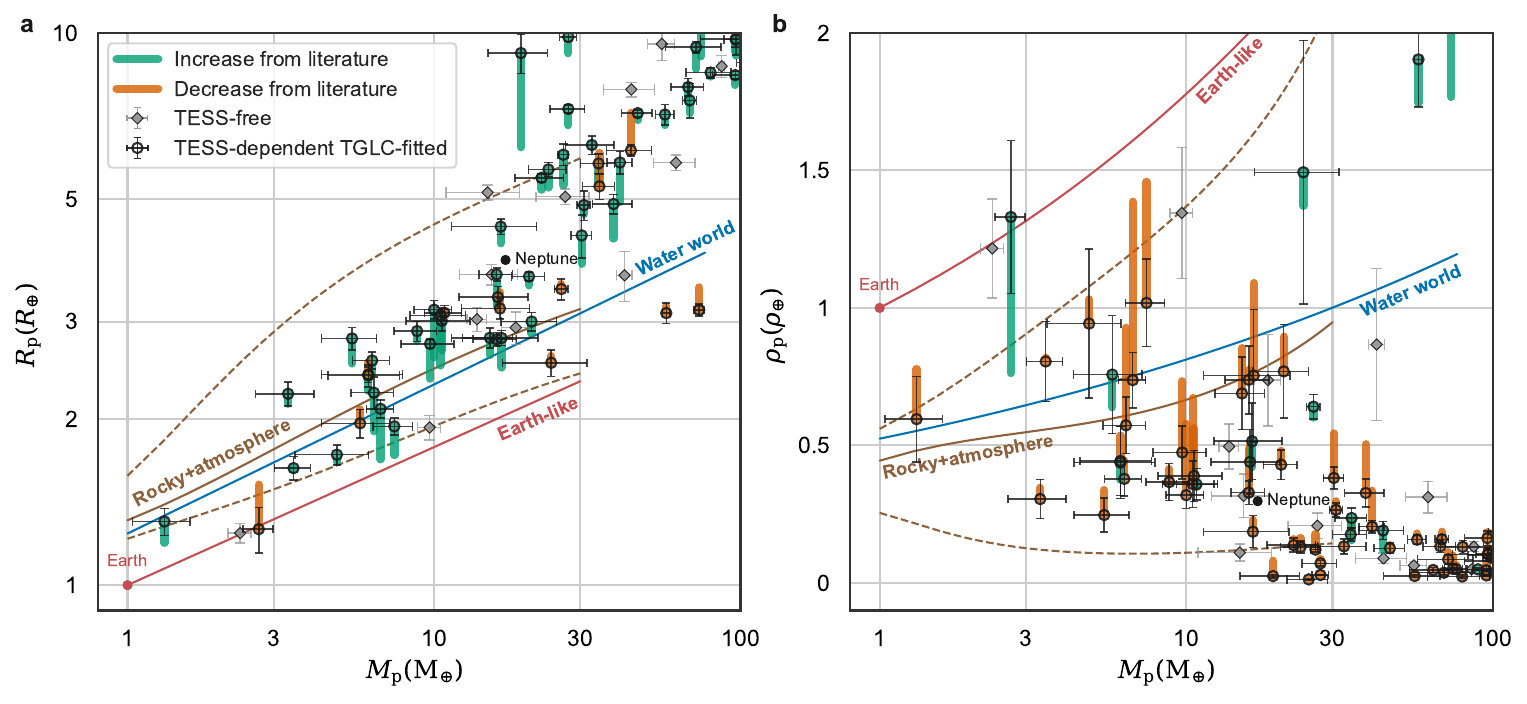}
    \caption{Planet mass-radius and mass-density distributions of literature and TGLC-fitted values. a, Mass-radius distribution of small TESS planets. b, Mass-density distribution of small TESS planets. Both panels include the high-precision planet sample with TESS-free literature values (gray diamond) and TESS-dependent TGLC-fitted values (black empty circle). The error bars represent $1\sigma$ Gaussian standard deviations. For each TESS-dependent TGLC-fitted value, a line connects it to the corresponding literature value, where the color of the line shows whether it increased (green) or decreased (orange) from the literature value. We also include theoretical models of planet compositions for the Earth-like (red), water worlds (blue), and rocky planets with atmospheres (brown). The last model accounts for the boil-off initial conditions (solid brown) and a range of agnostic initial conditions (region between dashed brown). Earth and Neptune are included for reference. }
    \label{fig:mass_density}
\end{figure*}

A systematic radius underestimation for hundreds of TESS exoplanets has profound implications for our general understanding of exoplanets. The immediate consequence is a biased distribution of these planets on the mass-radius ($M$-$R$) diagram. For smaller planets below $100\,\text{M}_\oplus$, we compare the distribution of the TESS-dependent planets in the $M$-$R$ diagram between literature and TGLC-fitted values. For this and subsequent analyses, we focus on the``high-precision planet" sample, which has mass measurement precisions better than 33\% and TGLC-fitted radius measurement precisions better than 20\%. This sample includes 26 TESS-free planets and 163 TESS-dependent planets. To calculate the TGLC-fitted radii, we first take the $p$ from all sectors of the planet. We then take the inverse-variance weighted mean of the $p$. We then multiply this mean with the stellar radius and propagate the errors. The TGLC-fitted densities are calculated with TGLC-fitted radii and literature planet masses. Stellar and planetary properties for each individual system are taken from the same publication as listed in Tables~\ref{tab:tess-free} and~\ref{tab:tess-dependent}, where we also listed $p_{\text{TGLC}}$ and $f_p$. 

We observe a population-wide shift towards larger planet radius (Figure~\ref{fig:mass_density}a). This shift directly challenges the inferred compositions of these planets from theoretical $M$-$R$ relations. Theoretical models have predicted distinct $M$-$R$ relations for Earth-like planets \citep{Owen2017}, water worlds \citep{Zeng2019}, and rocky planets with atmospheres \citep{Rogers2023}. For Earth-like planets, we use a power-law relation \citep{Owen2017} for planet mass ($M_{\text{p}}$) less than $30 \text{M}_{\oplus}$. We also model the water worlds \citep{Zeng2019} with $1.24 \text{R}_{\oplus} < R_{\text{p}} < 4 \text{R}_{\oplus}$. We also calculated the photoevaporation model of rocky planets with atmospheres ($M_{\text{p}} < 30 \text{M}_{\oplus}$) in boil-off and agnostic initial conditions \citep{Rogers2023}. Our corrected TGLC-fitted radii shift TESS-dependent planets by amounts comparable to the differences between these predicted compositions, altering our interpretation of their compositions. While the $M$-$R$ relation alone may not unambiguously distinguish these populations \citep{Rogers2023}, the radius bias we identify undermines compositional inferences for a significant fraction of the TESS exoplanets. We also observe a small population of planets with overestimated radii in the literature, suggesting that the contamination in the TESS photometry used by these studies may have been overcorrected.

Furthermore, corrected compositions could systematically alter exoplanet demographic studies: although our sample is not corrected for observational biases inherent to occurrence rate calculations, the widespread underestimation of radii would misrepresent the apparent prevalence of rocky versus gaseous planets, potentially reshaping our understanding of planetary architectures. The shifted radii also influence the distribution of TESS planets near the radius valley \citep{Owen2013}, but the scarcity of planets in this regime within our sample limits the ability to detect this shift in the TESS population alone. TESS exoplanets are often selected for atmospheric characterization using transmission spectroscopy with the James Webb Space Telescope \citep[JWST;][]{Gardner2006}. The larger TGLC-fitted radii of the TESS-dependent planets would also increase the transmission spectroscopy metric  \citep[TSM;][]{Kempton2018}, thus improving their predicted quality of observations and elevating their ranking as high-priority targets. 

The underestimated radii also strongly impact inferred planet densities. If the planets are larger than initially estimated by $\sim 6\%$, then their bulk densities would be correspondingly lower by $\sim 20\%$ (Figure~\ref{fig:mass_density}b). This affects the densest planets by the largest amount: all three TESS-dependent planets (TOI-1078, TOI-1235, TOI-4524) in our sample previously considered as near Earth-like are now more preferably described as rocky planets with atmospheres. This potentially suggests that these planets may not be as iron-rich as the Earth, and the distribution of TESS exoplanets near Earth-like composition may be sparser than initially suggested. 

\subsection{Water worlds in the TESS sample?}

\begin{figure*}[ht!]
    \centering
    \includegraphics[width=\textwidth]{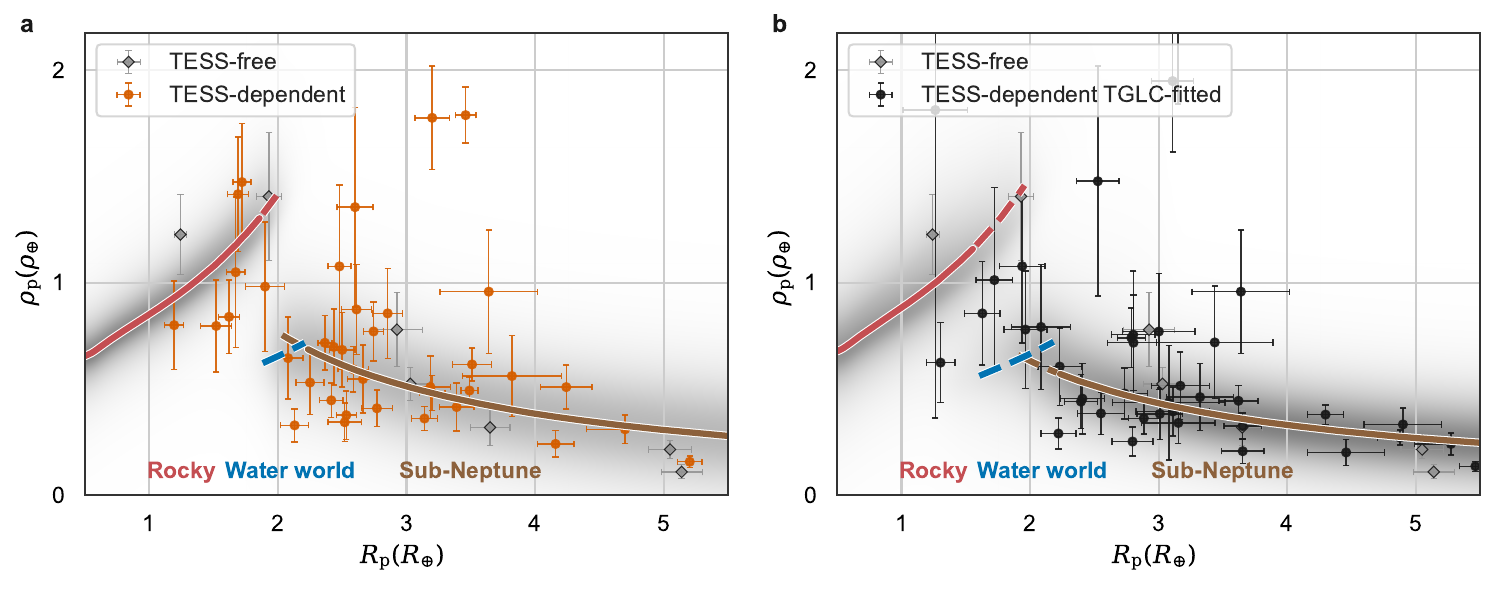}
    \caption{Probabilistic radius-density relations for small TESS planets. a, Probabilistic radius-density relations fitted to TESS-free and TESS-dependent literature values. b, Probabilistic radius-density relations fitted to TESS-free and TESS-dependent TGLC-fitted values. The figure includes the high-precision planet sample with TESS-free literature values (gray), TESS-dependent literature values (orange), and TESS-dependent TGLC-fitted values (black). The error bars represent $1\sigma$ Gaussian standard deviations. Probability density maps (light gray) are shown together with the most probable models. From small to large radii, the line components correspond to three planet compositions: rocky planets (red), water worlds (blue), and sub-Neptunes (brown). Dashed lines indicate regions where planetary populations overlap in radius space. TGLC-fitted radii for TESS planets yield a broader and more populated water world regime, suggesting a modestly stronger statistical preference for their existence compared to the model using literature values. }
    \label{fig:density}
\end{figure*}
The measured densities of small exoplanets have also been argued to separate rocky and water worlds around M dwarf stars \citep{Luque2022}. Although we lack M dwarf rocky planets in our sample to directly see the effect of the reduced densities, we can use the TESS exoplanets of all spectral types to show the preferred distribution of different populations of small planets. We use the \texttt{SPRIGHT} package \citep{Parviainen2024} to model the probabilistic mass–radius relation, which describes the combined probability distribution of planetary radius and bulk density as the mean of a three-component mixture model, separating planets into rocky, water world, and sub-Neptune populations. We apply this model to our high-precision planet sample to test whether a standalone population of water worlds is supported between rocky planets and sub-Neptunes. Only planets with radius smaller than $5.5 \,\text{R}_\oplus$ are used by the model. We set up two numerical radius–density probability models using the \texttt{RMEstimator} class: one with literature values for both TESS-free and TESS-dependent planets, and another where the radii of TESS-dependent planets are replaced with TGLC-fitted radii. We first apply the model directly to the selected high-precision planets using literature values, including both TESS-free and TESS-dependent planets (Figure~\ref{fig:density}a). The results show a narrow distribution of water worlds, suggesting that this population is statistically disfavored. However, if we replace the radii of the TESS-dependent planets with TGLC-fitted values, planets in the rocky planet regime move towards the water world regime (Figure~\ref{fig:density}b). As a result, the water world component extends further in the radius space, indicating a somewhat stronger preference towards the existence of water worlds.

\subsection{Recalibrating the empirical TESS exoplanet mass-radius relation}


Extending across the full range of planetary radii, this observational bias also directly impacts the observed $M$-$R$ relations. To demonstrate the scale of the impact, we compare the inferred $M$-$R$ relations using TESS literature radii and TGLC-fitted radii to a literature model \citep{Muller2024}, which is fitted using the PlanetS catalog of high-precision planets containing $\lesssim 20\%$ TESS targets. We selected the same high-precision planets as above and adopted a piecewise power-law relation with two breakpoints \citep{Muller2024}. Since we have sparse coverage for small planets, the original model with movable breakpoints cannot converge. We thus omit fitting for the small planets, and use a model with a single breakpoint at $127 \text{M}_\oplus$ according to the literature model \citep{Muller2024} and focus our comparison to the shifts in radius (Appendix~\ref{app:mr_relation}). 

We first fit a model with both TESS-free and TESS-dependent planets using literature planet masses and radii. The radii of the TESS-dependent planets are then replaced with TGLC-fitted radii and fitted to another model. The model using TGLC-fitted radii has a better agreement to the literature model (Appendix~\ref{app:mr_relation}), especially for the intermediate-mass planets \citep{Muller2024}. This result suggests that using the TESS literature radii leads to a somewhat biased $M$-$R$ relation, whereas applying TGLC-fitted radii yields a relation more consistent with the literature model dominated by unbiased measurements. For giant planets, a large dispersion remains among literature relations \citep{Muller2024,Edmondson2023,Hatzes2015,Chen2017}, but those relations encompass the results of our two models. While some newer results \citep{Muller2024,Edmondson2023} that include a small number of planets with TESS-influenced radii might require recalibration, the higher precision cut represented by these samples may reduce the effect. We conclude the TGLC-fitted radii shift the TESS population towards better agreement with the observed $M$-$R$ relations fitted to a high-precision catalog. 




\section{Synthesis and Path Forward} \label{sec:summary}
Our study reveals that unresolved blending in TESS photometry may be responsible for a systematic $\sim 6\%$ radius underestimation for 228 planets, skewing their inferred densities and compositions. This bias likely extends to many other TESS planets beyond our sample, suggesting a broader need for correction. Correction of these radius measurements would shift these planets toward lower densities, challenging previous claims of Earth-like planets, while enhancing the viability of some targets for JWST atmospheric studies. These revisions can be used to recalibrate demographic models and planet formation theories, emphasizing the need for deblended photometry in future studies. As precise measurements of planet radii refine population-level trends—in particular the radius valley and mass-radius relation—they will test the universality of planetary architectures, thus advancing our understanding of exoplanet diversity and evolution.

\section{Data Availability} \label{sec:data}
The TGLC light curves are publicly available through the Mikulski Archive for Space Telescopes (MAST) High-Level Science Products (HLSP) archive at \url{https://archive.stsci.edu/hlsp/tglc} \citep{tglc_hlsp}. The photometric fitting script, plotting scripts, posterior samples, and machine-readable versions of Tables~\ref{tab:tess-free},~\ref{tab:tess-dependent}, and~\ref{tab:kepler} are all available on Zenodo at \url{https://doi.org/10.5281/zenodo.15693253} \citep{Han2025}. We also include a \texttt{README} file in the Zenodo repository that explains how to reproduce the results using these data.

\begin{acknowledgments}
TH gratefully acknowledges insightful discussions with Simon Müller regarding the $M$–$R$ relation. TH also thanks Andrew Vanderburg for valuable input on the methods and the presentation of the results. 

This work was supported in part by NASA XRP grant 80NSSC23K0263.
\end{acknowledgments}

\begin{contribution}

TH developed TGLC, collected data, performed transit fitting and analysis, created figures, and wrote most of the manuscript. TH and PR analyzed TGLC radius measurements, and realzied the existence of a radius bias. PR also guided the experiment's development. TB, as TGLC's P.I., provided insights into statistical methods. SK and CC contributed to exoplanet fitting scripts and experimental design. AS and GR assisted in understanding literature methods and TESS mission characteristics. GR suggested Kepler planet comparisons and provided insights on population shifts. CB contributed to exoplanet fitting.
\end{contribution}

\appendix

\section{Exoplanet categorization} \label{app:categorization}
We found that methods for measuring exoplanet radii in the TESS literature fell into four categories:
\begin{enumerate}[label=\emph{Category \arabic*}:, leftmargin=*]
    \item The TESS light curve is modeled together with ground-based photometry, using a wide uniform prior on the dilution factor.
    \item The TESS light curve is modeled together with ground-based photometry, using a Gaussian or narrow uniform prior on the dilution factor.
    \item The TESS light curve is directly used in the fit to derive the planet radius, either alone or alongside ground-based photometry, without explicitly correcting for dilution.
    \item A fixed dilution factor is applied to the TESS light curve prior to fitting, and the corrected data is then used with or without ground-based photometry to derive the planet radius.
\end{enumerate}
\emph{Category 1} estimates a radius that is only determined by the ground-based photometry, which is usually much less diluted than TESS and gives more accurate transit depth measurements. The last three categories, on the other hand, estimate the radius that is completely/partially affected by the transit depth of the chosen TESS photometry pipeline. We categorize \emph{Category 1} as ``TESS-free" (Table~\ref{tab:tess-free}) and the last three categories together as ``TESS-dependent" (Table~\ref{tab:tess-dependent}). Our planet sample uses multiple TESS photometry products referenced in the literature, e.g., SPOC \citep{Jenkins2016}, TESS-SPOC \citep{Caldwell2020}, QLP \citep{Huang2020, Kunimoto2021}, $\texttt{eleanor}$ \citep{Feinstein2019}, TGLC \citep{Han2023}, $\texttt{DIAmante}$ \citep{Montalto2020}, PATHOS \citep{Nardiello2020}, and TASOC \citep{Lund2017}. Some studies also take the planet radii from the TOI catalog \citep{Guerrero2021} or use customized light curves.  

\section{Photometric fit priors and exoplanet sample used in this study}

\begin{longtable}{ccc} 
\caption{Priors used by all TGLC planet fits} \label{tab:prior} \\
\hline
Parameter & description & prior \\
\hline
\endfirsthead
\hline
Parameter & description & prior \\
\hline
\endhead
\hline\endfoot
$P$ & Orbital period (days) & $\mathcal{N}(P_{\text{lit}}, 0.1)$ \\
$T_C$ & Transit midpoint (BJD$_{\text{TDB}}$) & $\mathcal{N}((T_C)_{\text{lit}}, 0.01)$ \\
$p$ & Planet-to-star radius ratio & $\mathcal{LN}(\ln p_{\text{lit}}, 1)$ \\
$b$ & Impact parameter & $\mathcal{U}(0,1)$ \\
$q_1$ & Quadratic limb-darkening parameter & $\mathcal{U}(0,1)$ \\
$q_2$ & Quadratic limb-darkening parameter & $\mathcal{U}(0,1)$ \\
$e$ & Eccentricity & $\mathcal{N}(e_{\text{lit}}, \sigma_{e_{\text{lit}}})$ \\
$\omega$ & Argument of periastron & $\mathcal{N}(\omega_{\text{lit}}, \sigma_{\omega_{\text{lit}}})$ \\
\hline
\multicolumn{3}{l}{\textbf{Legend:}} \\
\multicolumn{3}{l}{Normal prior: $\mathcal{N}$(mean, standard deviation)} \\
\multicolumn{3}{l}{Log-normal prior: $\mathcal{LN}$(mean, standard deviation)} \\
\multicolumn{3}{l}{Uniform prior: $\mathcal{U}$(lower, upper)} \\
\multicolumn{3}{l}{$P_{\text{lit}}, (T_C)_{\text{lit}},$ and $p_{\text{lit}}$ correspond to the literature values. $e_{\text{lit}}$ and $\omega_{\text{lit}}$ are only applied to systems with $e_{\text{lit}} \geq 3 \sigma_{e_{\text{lit}}}$.} \\
\end{longtable}

\begin{longtable}{ccccc}
\caption{29 TESS-free planets and radius measurement methods} \label{tab:tess-free} \\
\hline
TIC & Photometry & Literature & $p_{\text{TGLC}}$ & $f_p$ \\
\hline
\endfirsthead
\hline
TIC & Photometry & Literature & $p_{\text{TGLC}}$ & $f_p$ \\
\hline
\endhead
\hline\endfoot
\\
\multicolumn{5}{l}{\emph{Category 1: TESS + ground-based, wide dilution prior}} \\
\\
16005254 &  TGLC & \cite{Han2024} & $0.168 \pm 0.004$ & 1.9\% $\pm$ 2.3\% \\
20182780 &  $\texttt{eleanor}$+ SPOC & \cite{Canas2023} & $0.170 \pm 0.004$ & 9.5\% $\pm$ 2.1\% \\
33595516 &  QLP & \cite{Armstrong2020} & $0.040 \pm 0.002$ & 14.9\% $\pm$ 4.0\% \\
44792534 & $\dagger$ & \cite{Sha2021} & $0.0487 \pm 0.0008$ & 4.9\% $\pm$ 1.5\% \\
119585136 &  TGLC & \cite{Kanodia2024} & $0.190 \pm 0.008$ & 6.8\% $\pm$ 3.9\% \\
144700903 &  SPOC & \cite{Kanodia2021} & $0.088 \pm 0.004$ & 0.7\% $\pm$ 4.3\% \\
150098860 &  SPOC & \cite{Hoyer2021} & $0.034 \pm 0.002$ & 3.4\% $\pm$ 4.8\% \\
154220877 &  TGLC & \cite{Kanodia2024} & $0.157 \pm 0.007$ & 2.4\% $\pm$ 4.5\% \\
179317684 & $\dagger$ & \cite{Kossakowski2019} & $0.0866 \pm 0.0002$ & -7.2\% $\pm$ 0.3\% \\
193641523 &  SPOC & \cite{Burt2020} & $0.043 \pm 0.002$ & 10.3\% $\pm$ 3.6\% \\
241249530 &  TESS-SPOC+ SPOC & \cite{Gupta2024} & $0.093 \pm 0.008$ & 6.4\% $\pm$ 7.8\% \\
243641947 &  TESS-SPOC & \cite{Hobson2023} & $0.284 \pm 0.008$ & 0.4\% $\pm$ 3.0\% \\
250111245 &  $\texttt{eleanor}$ & \cite{Canas2023} & $0.239 \pm 0.009$ & 12.2\% $\pm$ 3.2\% \\
259172249 &  TGLC & \cite{Kanodia2024} & $0.134 \pm 0.002$ & 4.9\% $\pm$ 1.7\% \\
271893367 &  SPOC & \cite{Kossakowski2019} & $0.0836 \pm 0.0002$ & -1.0\% $\pm$ 0.3\% \\
285048486 &  SPOC & \cite{Kanodia2020} & $0.076 \pm 0.002$ & 2.9\% $\pm$ 2.7\% \\
335590096 &  ground-only & \cite{Triaud2023} & $0.238 \pm 0.004$ & 6.3\% $\pm$ 1.7\% \\
376524552 &  SPOC & \cite{Rodriguez2023} & $0.146 \pm 0.002$ & 8.7\% $\pm$ 1.1\% \\
388076422 &  $\texttt{eleanor}$ & \cite{Kanodia2024} & $0.24 \pm 0.02$ & 6.9\% $\pm$ 8.7\% \\
394050135 & $\dagger$+ SPOC & \cite{Rodriguez2023} & $0.0786 \pm 0.0005$ & 6.4\% $\pm$ 0.6\% \\
396562848 &  SPOC & \cite{Soto2021} & $0.045 \pm 0.007$ & -3.5\% $\pm$ 16.7\% \\
409794137 &  QLP & \cite{Rodriguez2021} & $0.104 \pm 0.001$ & -0.4\% $\pm$ 1.2\% \\
419411415 &  $\texttt{eleanor}$ & \cite{Kanodia2023} & $0.263 \pm 0.005$ & -2.8\% $\pm$ 2.0\% \\
428787891 & $\dagger$ & \cite{Rodriguez2023} & $0.113 \pm 0.002$ & 5.2\% $\pm$ 1.7\% \\
445751830 &  $\texttt{eleanor}$ & \cite{Kanodia2022} & $0.191 \pm 0.005$ & 7.0\% $\pm$ 2.4\% \\
447061717 &  SPOC & \cite{Burt2021} & $0.075 \pm 0.002$ & 6.7\% $\pm$ 2.1\% \\
458419328 &  TESS-SPOC & \cite{Powers2023} & $0.102 \pm 0.002$ & 7.4\% $\pm$ 2.0\% \\
460984940 &  SPOC & \cite{Hua2023} & $0.021 \pm 0.001$ & 5.2\% $\pm$ 5.1\% \\
464300749 &  QLP$^a$ & \cite{Dong2021} & $0.067 \pm 0.001$ & 4.3\% $\pm$ 2.1\% \\
\hline
\multicolumn{5}{l}{\textbf{Legend:}} \\
\multicolumn{5}{l}{All TESS-free planets as defined fall into \emph{Category 1} in Appendix~\ref{app:categorization}.} \\
\multicolumn{5}{l}{TIC: TESS input catalog ID of the exoplanet host star \citep{Stassun2018}; Photometry: the reduction} \\
\multicolumn{5}{l}{method used in the literature for the planet radius fit; Literature: The reference literature selected for the planet;} \\
\multicolumn{5}{l}{$p_{\text{TGLC}}$: planet-to-star radius ratio measured with TGLC; $f_p$: fractional difference in planet-to-star radius} \\
\multicolumn{5}{l}{ ratios as defined in Equation~\ref{eq:1}.} \\
\multicolumn{5}{l}{\textbf{Footnotes:}} \\
\multicolumn{5}{l}{$\dagger$ An unnamed customized light curve extraction method is described in the according literature. } \\
\multicolumn{5}{l}{$^a$ The literature reported planet radius does not include a dilution factor, but one of their transit-only models } \\
\multicolumn{5}{l}{~~include a free dilution factor. We adopted radius ratio from the latter model. } \\
\end{longtable}

\begin{longtable}{ccccc} 
\caption{242 TESS-dependent planets and radius measurement methods} \label{tab:tess-dependent}\\
\hline
TIC & Photometry & Literature & $p_{\text{TGLC}}$ & $f_p$ \\
\hline
\endfirsthead
\hline
TIC & Photometry & Literature & $p_{\text{TGLC}}$ & $f_p$ \\
\hline
\endhead
\hline\endfoot
\\
\multicolumn{5}{l}{\emph{Category 2: TESS + ground-based, narrow dilution prior}} \\
\\
7548817 &  QLP+ SPOC & \cite{Yee2023} & $0.094 \pm 0.001$ & 4.1\% $\pm$ 1.0\% \\
8599009 &  QLP & \cite{Yee2023} & $0.125 \pm 0.006$ & 8.3\% $\pm$ 4.8\% \\
23769326 &  TESS-SPOC+ SPOC & \cite{Schulte2024} & $0.090 \pm 0.003$ & 5.0\% $\pm$ 3.4\% \\
39414571 &  QLP & \cite{Yee2023} & $0.097 \pm 0.001$ & 8.6\% $\pm$ 1.3\% \\
58825110 &  TESS-SPOC & \cite{Schulte2024} & $0.054 \pm 0.001$ & -1.0\% $\pm$ 2.5\% \\
66561343 &  $\texttt{tesseract}$ & \cite{Brahm2020} & $0.0891 \pm 0.0009$ & 11.2\% $\pm$ 0.9\% \\
68007716 &  QLP & \cite{Yee2023} & $0.066 \pm 0.001$ & 2.1\% $\pm$ 2.0\% \\
69356857 &  QLP & \cite{Schulte2024} & $0.073 \pm 0.001$ & 10.2\% $\pm$ 1.6\% \\
70524163 &  SPOC & \cite{Yee2022} & $0.057 \pm 0.001$ & 5.5\% $\pm$ 1.9\% \\
95660472 &  QLP+ SPOC & \cite{Yee2023} & $0.083 \pm 0.001$ & 5.9\% $\pm$ 1.4\% \\
100389539 &  QLP & \cite{Yee2023} & $0.086 \pm 0.001$ & 6.2\% $\pm$ 1.2\% \\
124379043 &  QLP & \cite{Yee2023} & $0.141 \pm 0.001$ & 5.6\% $\pm$ 0.9\% \\
139375960 &  QLP & \cite{Rodriguez2021} & $0.048 \pm 0.004$ & -20.5\% $\pm$ 8.8\% \\
147977348 &  QLP & \cite{Rodriguez2021} & $0.087 \pm 0.001$ & -0.6\% $\pm$ 1.2\% \\
151483286 &  QLP & \cite{Yee2023} & $0.115 \pm 0.002$ & 0.8\% $\pm$ 1.6\% \\
154293917 &  QLP & \cite{Yee2023} & $0.080 \pm 0.001$ & 6.0\% $\pm$ 1.7\% \\
156648452 &  QLP+ SPOC & \cite{Yee2023} & $0.097 \pm 0.003$ & 3.4\% $\pm$ 2.5\% \\
165464482 &  TESS-SPOC+ SPOC & \cite{Schulte2024} & $0.099 \pm 0.002$ & 6.2\% $\pm$ 1.5\% \\
166184428 & $\dagger$ & \cite{Timmermans2024} & $0.065 \pm 0.004$ & 8.6\% $\pm$ 5.8\% \\
178162579 &  QLP & \cite{Yee2023} & $0.097 \pm 0.002$ & 3.8\% $\pm$ 1.9\% \\
178709444 &  SPOC & \cite{Hartman2024} & $0.183 \pm 0.006$ & 1.6\% $\pm$ 3.0\% \\
194795551 &  QLP & \cite{Yee2022} & $0.125 \pm 0.002$ & 0.4\% $\pm$ 1.5\% \\
198008005 &  QLP+ SPOC & \cite{Hagelberg2023} & $0.106 \pm 0.002$ & 7.1\% $\pm$ 1.7\% \\
207110080 & $\dagger$+SPOC & \cite{Ikwut-Ukwa2022} & $0.09 \pm 0.01$ & 17.7\% $\pm$ 12.8\% \\
209459275 & $\dagger$+SPOC & \cite{Ikwut-Ukwa2022} & $0.0961 \pm 0.0008$ & 5.3\% $\pm$ 0.8\% \\
239816546 &  SPOC & \cite{Yee2022} & $0.121 \pm 0.002$ & 5.6\% $\pm$ 1.2\% \\
240823272 &  SPOC & \cite{Yee2022} & $0.149 \pm 0.002$ & 1.9\% $\pm$ 1.5\% \\
258920431 &  SPOC & \cite{Yee2022} & $0.0602 \pm 0.0006$ & 1.4\% $\pm$ 1.0\% \\
263179590 &  SPOC & \cite{Frame2023} & $0.038 \pm 0.002$ & -14.9\% $\pm$ 5.0\% \\
268301217 &  QLP+ SPOC & \cite{Yee2023} & $0.151 \pm 0.007$ & 21.7\% $\pm$ 3.5\% \\
269333648 &  $\texttt{tesseract}$ & \cite{Jones2024} & $0.059 \pm 0.002$ & -6.1\% $\pm$ 2.7\% \\
279947414 &  QLP & \cite{Yee2023} & $0.148 \pm 0.001$ & 4.3\% $\pm$ 1.0\% \\
280655495 &  QLP & \cite{Yee2023} & $0.079 \pm 0.001$ & -0.2\% $\pm$ 1.6\% \\
281408474 &  QLP & \cite{Rodriguez2021} & $0.081 \pm 0.001$ & -0.5\% $\pm$ 1.5\% \\
306648160 &  TESS-SPOC & \cite{Schulte2024} & $0.098 \pm 0.001$ & 5.6\% $\pm$ 1.2\% \\
310002617 &  QLP+ SPOC & \cite{Yee2023} & $0.1150 \pm 0.0007$ & 6.4\% $\pm$ 0.6\% \\
336128819 &  SPOC & \cite{Gan2022} & $0.060 \pm 0.003$ & 1.9\% $\pm$ 4.3\% \\
361343239 &  QLP & \cite{Yee2023} & $0.111 \pm 0.001$ & -1.1\% $\pm$ 1.3\% \\
375506058 &  SPOC & \cite{Addison2021} & $0.084 \pm 0.002$ & 5.5\% $\pm$ 1.8\% \\
394722182 &  SPOC & \cite{Yee2022} & $0.0874 \pm 0.0008$ & 4.2\% $\pm$ 0.9\% \\
395171208 &  QLP & \cite{Rodriguez2021} & $0.071 \pm 0.002$ & -5.4\% $\pm$ 3.5\% \\
401125028 &  TESS-SPOC+ SPOC & \cite{Schulte2024} & $0.125 \pm 0.004$ & 6.4\% $\pm$ 3.0\% \\
417646390 &  SPOC & \cite{Yee2022} & $0.092 \pm 0.001$ & 6.1\% $\pm$ 1.0\% \\
439366538 &  QLP & \cite{Hacker2024} & $0.09 \pm 0.01$ & 4.3\% $\pm$ 12.3\% \\
446549906 &  QLP & \cite{Schulte2024} & $0.141 \pm 0.003$ & 5.2\% $\pm$ 1.9\% \\
452006073 &  QLP & \cite{Hacker2024} & $0.053 \pm 0.002$ & 5.2\% $\pm$ 3.5\% \\
454248975 &  QLP & \cite{Yee2023} & $0.0951 \pm 0.0008$ & 5.1\% $\pm$ 0.8\% \\
470171739 &  TESS-SPOC+ SPOC & \cite{Schulte2024} & $0.0983 \pm 0.0006$ & 6.3\% $\pm$ 0.5\% \\
\hline
\\
\multicolumn{5}{l}{\emph{Category 3: no dilution correction}} \\
\\
1167538 &  SPOC & \cite{Gill2024} & $0.094 \pm 0.001$ & 7.3\% $\pm$ 1.0\% \\
4672985 &  $\texttt{tesseract}$ & \cite{Jones2024} & $0.10 \pm 0.01$ & 12.0\% $\pm$ 10.7\% \\
9348006 &  SPOC & \cite{Polanski2024} & $0.093 \pm 0.002$ & 4.0\% $\pm$ 1.8\% \\
10837041 & $\dagger$ SPOC$^a$ & \cite{Giacalone2022} & $0.023 \pm 0.002$ & -0.5\% $\pm$ 9.4\% \\
12421862 &  SPOC & \cite{Oddo2023} & $0.033 \pm 0.003$ & 9.6\% $\pm$ 9.5\% \\
22233480 &  SPOC & \cite{Goffo2024} & $0.069 \pm 0.002$ & 9.9\% $\pm$ 2.7\% \\
24358417 &  TESS-SPOC & \cite{Kabath2022} & $0.105 \pm 0.002$ & 6.9\% $\pm$ 1.6\% \\
29960110 &  SPOC & \cite{Kossakowski2021} & $0.046 \pm 0.003$ & 5.3\% $\pm$ 6.1\% \\
34077285 &  TOI & \cite{Hord2024} & $0.026 \pm 0.002$ & ... \\
35009898 &  TOI & \cite{Hord2024} & $0.052 \pm 0.003$ & ... \\
37770169 &  SPOC & \cite{Oddo2023} & $0.051 \pm 0.002$ & 6.5\% $\pm$ 3.3\% \\
49254857 &  TESS-SPOC & \cite{Mantovan2024} & $0.131 \pm 0.005$ & 9.9\% $\pm$ 3.1\% \\
49428710 &  SPOC & \cite{Mantovan2022} & $0.040 \pm 0.002$ & ... \\
54002556 & $\dagger$ & \cite{Gill2020} & $0.113 \pm 0.005$ & 10.6\% $\pm$ 3.9\% \\
58542531 &  TOI & \cite{Hord2024} & $0.031 \pm 0.002$ & ... \\
62483237 &  SPOC & \cite{Mistry2023} & $0.035 \pm 0.003$ & 8.9\% $\pm$ 6.9\% \\
70899085 &  SPOC & \cite{Dreizler2020} & $0.077 \pm 0.002$ & 3.6\% $\pm$ 2.3\% \\
73540072 &  SPOC & \cite{Naponiello2023} & $0.036 \pm 0.003$ & -9.8\% $\pm$ 9.4\% \\
76923707 &  SPOC & \cite{Mistry2023b} & $0.088 \pm 0.002$ & 2.8\% $\pm$ 1.8\% \\
83092282 &  SPOC & \cite{Psaridi2023} & $0.0631 \pm 0.0006$ & 5.2\% $\pm$ 0.9\% \\
91987762 &  SPOC & \cite{Polanski2024} & $0.059 \pm 0.001$ & 8.9\% $\pm$ 1.7\% \\
97568467 & $\dagger$+ SPOC  & \cite{Rodriguez2023} & $0.045 \pm 0.001$ & 3.2\% $\pm$ 3.0\% \\
97766057 &  SPOC & \cite{Saunders2024} & $0.077 \pm 0.002$ & -0.7\% $\pm$ 2.5\% \\
99869022 &  TESS-SPOC+ SPOC & \cite{Serrano-Bell2024} & $0.072 \pm 0.001$ & 8.2\% $\pm$ 1.6\% \\
103633434 &  SPOC & \cite{Luque2022} & $0.031 \pm 0.003$ & 19.0\% $\pm$ 8.5\% \\
118327550 &  SPOC & \cite{Castro-Gonzalez2023} & $0.027 \pm 0.005$ & -20.4\% $\pm$ 22.9\% \\
119584412 &  SPOC & \cite{Mallorquin2023} & $0.033 \pm 0.002$ & -5.9\% $\pm$ 7.7\% \\
120826158 &  TOI & \cite{Hord2024} & $0.030 \pm 0.001$ & ... \\
124573851 &  SPOC & \cite{Polanski2024} & $0.029 \pm 0.003$ & 18.0\% $\pm$ 7.3\% \\
126606859 &  SPOC & \cite{Esparza-Borges2022} & $0.058 \pm 0.003$ & 0.6\% $\pm$ 5.5\% \\
130924120 &  SPOC & \cite{Alqasim2024} & $0.036 \pm 0.005$ & 18.8\% $\pm$ 11.1\% \\
139285832 &  SPOC & \cite{Osborn2023} & $0.033 \pm 0.001$ & -2.9\% $\pm$ 4.5\% \\
140691463 &  QLP+ SPOC & \cite{Nielsen2020} & $0.1229 \pm 0.0005$ & 7.8\% $\pm$ 0.4\% \\
142381532 &  SPOC & \cite{Polanski2024} & $0.091 \pm 0.002$ & 5.5\% $\pm$ 1.7\% \\
142387023 &  SPOC & \cite{Polanski2024} & $0.031 \pm 0.001$ & 1.6\% $\pm$ 3.6\% \\
142394656 &  SPOC & \cite{Subjak2022} & $0.0969 \pm 0.0005$ & 6.4\% $\pm$ 0.5\% \\
142937186 & $\dagger$ SPOC$^a$ & \cite{Giacalone2022} & $0.024 \pm 0.002$ & -6.9\% $\pm$ 10.7\% \\
147950620 &  QLCP & \cite{Wang2023} & $0.0888 \pm 0.0009$ & 6.4\% $\pm$ 1.0\% \\
148673433 &  SPOC & \cite{Mistry2023} & $0.043 \pm 0.002$ & 8.3\% $\pm$ 4.9\% \\
151825527 &  SPOC & \cite{Mistry2023} & $0.092 \pm 0.002$ & 3.7\% $\pm$ 1.7\% \\
153065527 &  TOI & \cite{Hord2024} & $0.047 \pm 0.003$ & ... \\
154089169 &  SPOC & \cite{Polanski2024} & $0.030 \pm 0.001$ & 1.4\% $\pm$ 4.3\% \\
154872375 &  SPOC & \cite{Mallorquin2024} & $0.0754 \pm 0.0005$ & 5.4\% $\pm$ 0.6\% \\
158002130 &  SPOC & \cite{Polanski2024} & $0.0404 \pm 0.0009$ & 7.1\% $\pm$ 2.1\% \\
158025009 &  SPOC & \cite{Deeg2023} & $0.019 \pm 0.001$ & 0.5\% $\pm$ 7.5\% \\
158241252 &  TOI & \cite{Hord2024} & $0.016 \pm 0.002$ & ... \\
158588995 &  SPOC & \cite{Murgas2021} & $0.124 \pm 0.002$ & 7.2\% $\pm$ 1.5\% \\
159418353 &  SPOC & \cite{Mistry2024} & $0.033 \pm 0.005$ & 37.7\% $\pm$ 9.2\% \\
159781361 &  SPOC & \cite{Polanski2024} & $0.030 \pm 0.001$ & 8.1\% $\pm$ 4.5\% \\
159873822 &  SPOC & \cite{Newton2022} & $0.036 \pm 0.002$ & 16.1\% $\pm$ 4.2\% \\
160390955 & $\dagger$ & \cite{Martioli2024} & $0.077 \pm 0.003$ & 11.9\% $\pm$ 2.9\% \\
163539739 &  SPOC & \cite{Artigau2021} & $0.16 \pm 0.01$ & -25.1\% $\pm$ 11.9\% \\
166648874 &  TOI & \cite{Hord2024} & $0.079 \pm 0.006$ & ... \\
169765334 &  SPOC & \cite{Polanski2024} & $0.035 \pm 0.003$ & 15.2\% $\pm$ 6.4\% \\
172370679 &  SPOC & \cite{Canas2020} & $0.178 \pm 0.004$ & 6.0\% $\pm$ 2.0\% \\
176956893 &  $\texttt{giants}$ & \cite{Saunders2022} & $0.038 \pm 0.001$ & 5.4\% $\pm$ 2.8\% \\
179034327 &  SPOC & \cite{Oddo2023} & $0.039 \pm 0.001$ & 15.7\% $\pm$ 3.2\% \\
183120439 &  QLP+ SPOC & \cite{Nielsen2020} & $0.085 \pm 0.004$ & -1.5\% $\pm$ 4.7\% \\
183985250 &  SPOC & \cite{Jenkins2020} & $0.06 \pm 0.01$ & 19.9\% $\pm$ 18.0\% \\
190496853 &  QLP+ SPOC & \cite{Psaridi2023} & $0.108 \pm 0.001$ & 6.6\% $\pm$ 1.2\% \\
192790476 &  SPOC & \cite{Mistry2023} & $0.044 \pm 0.001$ & 8.2\% $\pm$ 2.5\% \\
198241702 &  SPOC & \cite{Polanski2024} & $0.027 \pm 0.002$ & 2.9\% $\pm$ 6.5\% \\
198356533 &   SPOC & \cite{Pidhorodetska2024} & $0.027 \pm 0.001$ & 34.0\% $\pm$ 2.7\% \\
198485881 &  SPOC & \cite{Schanche2022} & $0.070 \pm 0.004$ & 7.6\% $\pm$ 5.4\% \\
199376584 & $\dagger$ & \cite{Dalba2020} & $0.0821 \pm 0.0007$ & 19.9\% $\pm$ 0.7\% \\
200723869 &  SPOC & \cite{Addison2021b} & $0.030 \pm 0.003$ & -16.9\% $\pm$ 10.8\% \\
207141131 & $\dagger$+ SPOC & \cite{Desidera2023} & $0.030 \pm 0.002$ & -2.9\% $\pm$ 5.9\% \\
219016883 &  SPOC & \cite{Mantovan2022} & $0.047 \pm 0.002$ & ... \\
219041246 &  SPOC & \cite{Ghachoui2024} & $0.059 \pm 0.005$ & 7.3\% $\pm$ 7.7\% \\
219344917 &  SPOC & \cite{Mistry2024} & $0.023 \pm 0.002$ & 7.3\% $\pm$ 7.5\% \\
219850915 &   SPOC & \cite{Polanski2024} & $0.032 \pm 0.001$ & 5.8\% $\pm$ 3.0\% \\
219854185 &  SPOC & \cite{Moutou2021} & $0.0815 \pm 0.0003$ & 6.8\% $\pm$ 0.3\% \\
219854519 &  $\texttt{giants}$+ SPOC & \cite{Grunblatt2023} & $0.037 \pm 0.001$ & -4.0\% $\pm$ 4.2\% \\
219857012 &  SPOC & \cite{Polanski2024} & $0.022 \pm 0.001$ & 13.5\% $\pm$ 5.5\% \\
220479565 &  SPOC & \cite{Cointepas2021} & $0.066 \pm 0.002$ & 4.0\% $\pm$ 2.8\% \\
224297258 &  SPOC & \cite{Polanski2024} & $0.033 \pm 0.002$ & 11.6\% $\pm$ 5.0\% \\
229510866 &  SPOC & \cite{Kabath2022} & $0.0817 \pm 0.0002$ & 2.6\% $\pm$ 0.2\% \\
229742722 &  SPOC & \cite{Dong2023} & $0.076 \pm 0.001$ & 13.7\% $\pm$ 1.2\% \\
230001847 &  $\texttt{giants}$ & \cite{Grunblatt2022} & $0.029 \pm 0.002$ & 2.1\% $\pm$ 5.2\% \\
231702397 &  SPOC & \cite{Waalkes2021} & $0.078 \pm 0.006$ & 4.5\% $\pm$ 6.8\% \\
232540264 &  SPOC & \cite{Polanski2024} & $0.021 \pm 0.001$ & -5.2\% $\pm$ 7.2\% \\
232612416 &  SPOC & \cite{Polanski2024} & $0.0769 \pm 0.0005$ & 6.5\% $\pm$ 0.5\% \\
232967440 &  SPOC & \cite{Yana-Galarza2024} & $0.0965 \pm 0.0008$ & 6.5\% $\pm$ 0.8\% \\
232976128 &  SPOC & \cite{Polanski2024} & $0.031 \pm 0.002$ & 2.2\% $\pm$ 6.1\% \\
232982558 &  SPOC & \cite{Polanski2024} & $0.028 \pm 0.002$ & 13.4\% $\pm$ 5.1\% \\
233087860 &  SPOC & \cite{Polanski2024} & $0.0315 \pm 0.0008$ & -0.5\% $\pm$ 2.5\% \\
237104103 &  SPOC & \cite{Moutou2021} & $0.0648 \pm 0.0003$ & 6.1\% $\pm$ 0.5\% \\
237232044 &  SPOC & \cite{Polanski2024} & $0.029 \pm 0.001$ & 2.9\% $\pm$ 4.7\% \\
237913194 &  $\texttt{tesseract}$ & \cite{Schlecker2020} & $0.110 \pm 0.008$ & 3.8\% $\pm$ 7.4\% \\
246965431 &  SPOC & \cite{Pelaez-Torres2024} & $0.066 \pm 0.004$ & 19.0\% $\pm$ 4.6\% \\
256722647 &  $\texttt{giants}$ & \cite{Grunblatt2022} & $0.060 \pm 0.004$ & -11.7\% $\pm$ 7.5\% \\
257060897 & $\dagger$ & \cite{Montalto2022} & $0.0897 \pm 0.0003$ & 6.2\% $\pm$ 0.3\% \\
257527578 &  QLP+ SPOC & \cite{Ulmer-Moll2022} & $0.064 \pm 0.003$ & 3.3\% $\pm$ 4.5\% \\
260004324 &  SPOC & \cite{Luque2022} & $0.020 \pm 0.003$ & -0.4\% $\pm$ 16.6\% \\
261867566 &  SPOC & \cite{Davis2020} & $0.132 \pm 0.002$ & 1.1\% $\pm$ 1.5\% \\
262530407 &  SPOC & \cite{Almenara2022b} & $0.039 \pm 0.002$ & 4.1\% $\pm$ 5.1\% \\
266980320 &  SPOC & \cite{Oddo2023} & $0.045 \pm 0.002$ & 16.7\% $\pm$ 4.1\% \\
267263253 &  SPOC & \cite{Jones2019} & $0.1010 \pm 0.0003$ & 6.1\% $\pm$ 0.3\% \\
268334473 &  SPOC & \cite{Polanski2024} & $0.031 \pm 0.001$ & -3.7\% $\pm$ 4.7\% \\
268532343 &  QLP & \cite{Carleo2024} & $0.055 \pm 0.002$ & -5.8\% $\pm$ 3.3\% \\
271169413 &  SPOC & \cite{Mistry2024} & $0.037 \pm 0.003$ & 5.2\% $\pm$ 7.7\% \\
271478281 &  SPOC & \cite{Mistry2023} & $0.031 \pm 0.002$ & 13.6\% $\pm$ 6.8\% \\
271971130 &  SPOC & \cite{Dransfield2024} & $0.076 \pm 0.007$ & 22.4\% $\pm$ 7.4\% \\
277634430 &  SPOC & \cite{Mistry2024} & $0.056 \pm 0.002$ & 3.3\% $\pm$ 3.4\% \\
280206394 &  SPOC & \cite{Jordan2020} & $0.099 \pm 0.001$ & 5.4\% $\pm$ 0.9\% \\
282485660 &  TESS-SPOC+ SPOC & \cite{Castro-Gonzalez2024} & $0.062 \pm 0.002$ & 0.2\% $\pm$ 3.1\% \\
286916251 &  SPOC & \cite{Polanski2024} & $0.029 \pm 0.005$ & 20.5\% $\pm$ 13.2\% \\
287080092 &  SPOC & \cite{Polanski2024} & $0.021 \pm 0.005$ & -0.2\% $\pm$ 26.0\% \\
287145649 &  QLP+ TESS-SPOC & \cite{Mantovan2024} & $0.123 \pm 0.002$ & 7.6\% $\pm$ 1.5\% \\
288735205 &  SPOC & \cite{Martin2021} & $0.1583 \pm 0.0007$ & 6.7\% $\pm$ 0.4\% \\
289164482 &  SPOC & \cite{Pelaez-Torres2024} & $0.027 \pm 0.005$ & -30.7\% $\pm$ 21.8\% \\
289580577 &  SPOC & \cite{Polanski2024} & $0.026 \pm 0.002$ & 11.6\% $\pm$ 5.1\% \\
290348383 &  SPOC & \cite{Barros2023} & $0.029 \pm 0.001$ & 6.1\% $\pm$ 4.0\% \\
293954617 &  SPOC & \cite{Orell-Miquel2023} & $0.024 \pm 0.001$ & -0.8\% $\pm$ 4.2\% \\
296739893 &  SPOC & \cite{Reefe2022} & $0.069 \pm 0.006$ & 9.2\% $\pm$ 7.9\% \\
303432813 &  $\texttt{DIAmante}$ & \cite{Montalto2024} & $0.043 \pm 0.002$ & 12.4\% $\pm$ 3.3\% \\
305424003 &  TOI & \cite{Hord2024} & $0.039 \pm 0.004$ & ... \\
305739565 &  QLP & \cite{Bouchy2024} & $0.089 \pm 0.002$ & 8.5\% $\pm$ 1.7\% \\
306955329 &  SPOC & \cite{Polanski2024} & $0.049 \pm 0.003$ & 6.7\% $\pm$ 6.1\% \\
307958020 &  SPOC & \cite{Eisner2024} & $0.021 \pm 0.004$ & -30.5\% $\pm$ 24.5\% \\
317548889 &  SPOC & \cite{Polanski2024} & $0.018 \pm 0.002$ & 4.9\% $\pm$ 8.6\% \\
318753380 &  SPOC & \cite{Mistry2023} & $0.035 \pm 0.002$ & 0.5\% $\pm$ 6.6\% \\
320004517 &  SPOC & \cite{Palatnick2021} & $0.035 \pm 0.001$ & 3.6\% $\pm$ 4.0\% \\
321669174 &  SPOC & \cite{Esparza-Borges2022} & $0.037 \pm 0.003$ & 6.4\% $\pm$ 7.7\% \\
321857016 &  SPOC & \cite{Yoshida2023} & $0.125 \pm 0.002$ & 5.2\% $\pm$ 1.2\% \\
322807371 &  SPOC & \cite{Battley2024} & $0.111 \pm 0.003$ & 5.9\% $\pm$ 2.9\% \\
328081248 &  SPOC & \cite{Mistry2024} & $0.048 \pm 0.003$ & 6.2\% $\pm$ 6.5\% \\
328934463 &  TESS-SPOC+ SPOC & \cite{Carleo2024} & $0.070 \pm 0.002$ & 8.1\% $\pm$ 2.0\% \\
332534326 &  TESS-SPOC & \cite{Mantovan2024} & $0.106 \pm 0.003$ & 5.1\% $\pm$ 2.5\% \\
333657795 &  SPOC & \cite{Murgas2022} & $0.016 \pm 0.001$ & 11.3\% $\pm$ 8.0\% \\
335630746 &  SPOC & \cite{Clark2023} & $0.0890 \pm 0.0009$ & 7.2\% $\pm$ 0.9\% \\
342642208 &  QLP+ SPOC & \cite{Knudstrup2022} & $0.073 \pm 0.001$ & 3.8\% $\pm$ 1.7\% \\
343628284 &  TESS-SPOC+ SPOC & \cite{Hori2024} & $0.077 \pm 0.008$ & 14.1\% $\pm$ 8.5\% \\
348755728 &  SPOC & \cite{Pelaez-Torres2024} & $0.111 \pm 0.005$ & 5.6\% $\pm$ 3.9\% \\
349095149 &  SPOC & \cite{Mann2023} & $0.020 \pm 0.007$ & -27.4\% $\pm$ 41.3\% \\
350153977 &  SPOC & \cite{Hawthorn2023} & $0.030 \pm 0.002$ & 4.1\% $\pm$ 6.9\% \\
350618622 &  SPOC & \cite{Hobson2021} & $0.0838 \pm 0.0007$ & 6.2\% $\pm$ 0.8\% \\
353475866 & $\dagger$ SPOC$^a$ & \cite{Giacalone2022} & $0.028 \pm 0.004$ & -0.7\% $\pm$ 15.6\% \\
358070912 & $\dagger$ +  SPOC & \cite{Nabbie2024} & $0.046 \pm 0.001$ & 11.1\% $\pm$ 2.0\% \\
362249359 & $\dagger$ SPOC$^a$ & \cite{Giacalone2022} & $0.019 \pm 0.003$ & -1.9\% $\pm$ 15.0\% \\
367858035 &  SPOC & \cite{Polanski2024} & $0.042 \pm 0.003$ & 5.9\% $\pm$ 6.2\% \\
368287008 &  SPOC & \cite{Jones2024b} & $0.087 \pm 0.005$ & -7.0\% $\pm$ 6.7\% \\
370133522 &  SPOC & \cite{Luque2022} & $0.031 \pm 0.002$ & 8.5\% $\pm$ 6.2\% \\
372172128 &  SPOC & \cite{Persson2022} & $0.030 \pm 0.004$ & -2.1\% $\pm$ 13.4\% \\
376637093 &  SPOC & \cite{Kabath2022} & $0.1257 \pm 0.0006$ & 2.4\% $\pm$ 0.5\% \\
380887434 &  TOI & \cite{Hord2024} & $0.018 \pm 0.001$ & ... \\
387690507 &  SPOC & \cite{Gan2022b} & $0.150 \pm 0.005$ & -5.4\% $\pm$ 3.2\% \\
391949880 &  TOI & \cite{Hord2024} & $0.026 \pm 0.002$ & ... \\
392476080 &  QLP & \cite{Wong2021} & $0.084 \pm 0.002$ & 3.2\% $\pm$ 2.1\% \\
393818343 &  SPOC & \cite{Sgro2024} & $0.110 \pm 0.002$ & 6.8\% $\pm$ 1.4\% \\
394137592 &  SPOC & \cite{Nielsen2019} & $0.040 \pm 0.003$ & -12.3\% $\pm$ 7.9\% \\
394357918 &  SPOC & \cite{Barkaoui2023} & $0.103 \pm 0.008$ & 8.0\% $\pm$ 6.9\% \\
394561119 &  QLP+ SPOC & \cite{Psaridi2022} & $0.0803 \pm 0.0004$ & 8.0\% $\pm$ 0.5\% \\
394918211 &  $\texttt{giants}$ & \cite{Pereira2024} & $0.032 \pm 0.002$ & -23.9\% $\pm$ 8.5\% \\
395393265 & $\dagger$ & \cite{Rodriguez2023} & $0.087 \pm 0.001$ & 5.7\% $\pm$ 1.1\% \\
404505029 &  SPOC & \cite{Wittenmyer2022} & $0.055 \pm 0.001$ & -1.2\% $\pm$ 2.3\% \\
404518509 &  SPOC & \cite{Nies2024} & $0.020 \pm 0.003$ & -15.7\% $\pm$ 14.9\% \\
407126408 &  SPOC & \cite{Mistry2023} & $0.031 \pm 0.003$ & 0.9\% $\pm$ 8.8\% \\
407591297 &  TOI & \cite{Hord2024} & $0.035 \pm 0.004$ & ... \\
408636441 &  SPOC & \cite{Espinoza2022} & $0.048 \pm 0.004$ & -0.9\% $\pm$ 8.9\% \\
410214986 &  SPOC & \cite{Newton2019} & $0.056 \pm 0.004$ & 3.9\% $\pm$ 6.8\% \\
417931607 &  SPOC & \cite{Polanski2024} & $0.025 \pm 0.002$ & 6.9\% $\pm$ 6.6\% \\
419523962 &  SPOC & \cite{Saunders2024} & $0.046 \pm 0.004$ & -22.1\% $\pm$ 10.8\% \\
420112589 &  SPOC & \cite{Cadieux2022} & $0.057 \pm 0.004$ & 2.9\% $\pm$ 7.3\% \\
422756130 &  SPOC & \cite{Cherubim2023} & $0.040 \pm 0.003$ & 14.8\% $\pm$ 6.0\% \\
428699140 &  SPOC & \cite{Mistry2023} & $0.049 \pm 0.002$ & -1.0\% $\pm$ 4.7\% \\
429358906 &  TESS-SPOC+ SPOC & \cite{Hori2024} & $0.066 \pm 0.006$ & 8.4\% $\pm$ 7.7\% \\
436873727 &  SPOC & \cite{Bieryla2024} & $0.057 \pm 0.001$ & 23.5\% $\pm$ 1.4\% \\
437011608 &  TOI & \cite{Hord2024} & $0.027 \pm 0.001$ & ... \\
437856897 &  SPOC & \cite{Khandelwal2023} & $0.039 \pm 0.001$ & -1.5\% $\pm$ 3.5\% \\
439456714 &  SPOC & \cite{Magliano2023} & $0.076 \pm 0.004$ & 11.9\% $\pm$ 4.7\% \\
441738827 &  SPOC & \cite{Barkaoui2023} & $0.07 \pm 0.02$ & 29.1\% $\pm$ 20.6\% \\
441765914 &  SPOC & \cite{Polanski2024} & $0.038 \pm 0.002$ & -4.7\% $\pm$ 4.3\% \\
445805961 &  SPOC & \cite{Konig2022} & $0.0523 \pm 0.0008$ & 4.8\% $\pm$ 1.5\% \\
445903569 &  QLP+ SPOC & \cite{Page2024} & $0.064 \pm 0.001$ & 14.6\% $\pm$ 1.9\% \\
456862677 &  PATHOS+ SPOC & \cite{Carleo2024b} & $0.141 \pm 0.002$ & 8.2\% $\pm$ 1.3\% \\
460205581 &  SPOC & \cite{Bouma2020} & $0.08 \pm 0.01$ & 3.1\% $\pm$ 12.1\% \\
464646604 &  SPOC & \cite{Zhou2022} & $0.028 \pm 0.007$ & 8.9\% $\pm$ 23.7\% \\
466206508 &  SPOC & \cite{Grieves2022} & $0.104 \pm 0.004$ & 5.9\% $\pm$ 3.5\% \\
468574941 &  SPOC & \cite{Kabath2022} & $0.124 \pm 0.002$ & 1.1\% $\pm$ 1.9\% \\
\hline
\\
\multicolumn{5}{l}{\emph{Category 4: fixed dilution factor}} \\
\\
124029677 &  QLP+ SPOC & \cite{Ulmer-Moll2022} & $0.083 \pm 0.001$ & 6.1\% $\pm$ 1.5\% \\
157698565 &  \texttt{tesseract} & \cite{Brahm2023} & $0.106 \pm 0.002$ & 1.7\% $\pm$ 2.2\% \\
201177276 & $\dagger$ + SPOC  & \cite{Bryant2024} & $0.176 \pm 0.004$ & 2.1\% $\pm$ 2.1\% \\
206541859 &  \texttt{tesseract} & \cite{Brahm2023} & $0.090 \pm 0.004$ & 11.9\% $\pm$ 3.7\% \\
218795833 &  SPOC & \cite{Parviainen2021} & $0.31 \pm 0.01$ & 1.1\% $\pm$ 3.1\% \\
293607057 &  QLP+ SPOC & \cite{Subjak2024} & $0.085 \pm 0.002$ & 0.7\% $\pm$ 2.6\% \\
360156606 &  SPOC & \cite{Mann2022} & $0.15 \pm 0.01$ & -7.3\% $\pm$ 8.0\% \\
382602147 &  QLP+ SPOC & \cite{Bryant2024} & $0.180 \pm 0.002$ & 4.1\% $\pm$ 1.1\% \\
393831507 &  QLP+ SPOC & \cite{Knudstrup2022} & $0.083 \pm 0.002$ & 6.7\% $\pm$ 2.5\% \\
\hline
\\
\multicolumn{5}{l}{\emph{Mixed Categories}} \\
\\
169249234 & $4.$ QLP+$3.$ SPOC & \cite{Eberhardt2023} & $0.107 \pm 0.001$ & 14.0\% $\pm$ 1.1\% \\
237922465 & $4.$ QLP+$3.$ SPOC & \cite{Eberhardt2023} & $0.085 \pm 0.001$ & 13.9\% $\pm$ 1.5\% \\
332558858 & $4.$ QLP+$3.$ SPOC & \cite{Eberhardt2023} & $0.091 \pm 0.004$ & 4.5\% $\pm$ 4.7\% \\
339672028 & $4.$ $\texttt{tesseract}$+$3.$ SPOC & \cite{Brahm2020} & $0.0663 \pm 0.0003$ & 7.6\% $\pm$ 0.5\% \\
365102760 & $3.$ $\texttt{giants}$+$2.$ SPOC & \cite{Grunblatt2024} & $0.027 \pm 0.004$ & 32.5\% $\pm$ 10.7\% \\
\hline
\multicolumn{5}{l}{\textbf{Legend:}} \\
\multicolumn{5}{l}{Same as Table~\ref{tab:tess-free}, but for \emph{Category 2}, \emph{3}, and \emph{4} in Appendix~\ref{app:categorization}. $f_p$ values marked with `...' indicate 14 planets } \\
\multicolumn{5}{l}{without published uncertainties and are excluded from our analysis (see Section~\ref{sec:bias}).} \\
\multicolumn{5}{l}{\textbf{Footnotes:}} \\
\multicolumn{5}{l}{$\dagger$ An unnamed customized light curve extraction method is described in the according literature. } \\
\multicolumn{5}{l}{$^a$ A customized light curve is made for photometry fitting, but it uses the SPOC crowding factor to decontaminate.} \\
\end{longtable}

\begin{longtable}{cccc} 
\caption{30 Kepler planets} \label{tab:kepler}\\
\hline
TIC & Literature & $p_{\text{TGLC}}$ & $f_p$ \\
\hline
\endfirsthead
\hline
TIC & Literature & $p_{\text{TGLC}}$ & $f_p$ \\
\hline
\endhead
\hline\endfoot
27318774 & \cite{Morton2016} & $0.120 \pm 0.005$ & -0.7\% $\pm$ 4.2\% \\
27774415 & \cite{Endl2011} & $0.101 \pm 0.002$ & 1.3\% $\pm$ 1.9\% \\
27916356 & \cite{Esteves2015} & $0.098 \pm 0.002$ & 1.8\% $\pm$ 1.9\% \\
27990610 & \cite{Morton2016} & $0.122 \pm 0.005$ & 0.1\% $\pm$ 4.0\% \\
63452790 & \cite{Gandolfi2015} & $0.123 \pm 0.006$ & -5.2\% $\pm$ 4.9\% \\
122441491 & \cite{Hebrard2014} & $0.113 \pm 0.005$ & -3.6\% $\pm$ 4.9\% \\
122596693 & \cite{Hebrard2014} & $0.138 \pm 0.004$ & -0.9\% $\pm$ 2.6\% \\
123233041 & \cite{Esteves2015} & $0.096 \pm 0.002$ & -2.3\% $\pm$ 1.8\% \\
123495874 & \cite{Endl2014} & $0.095 \pm 0.002$ & 0.1\% $\pm$ 1.6\% \\
137683938 & \cite{Morton2016} & $0.105 \pm 0.009$ & -6.3\% $\pm$ 9.6\% \\
137899948 & \cite{Johnson2012} & $0.180 \pm 0.006$ & 0.5\% $\pm$ 3.1\% \\
138430864 & \cite{Morton2016} & $0.11 \pm 0.02$ & 11.5\% $\pm$ 13.3\% \\
158170594 & \cite{Esteves2015} & $0.107 \pm 0.003$ & 0.2\% $\pm$ 2.4\% \\
158388163 & \cite{Morton2016} & $0.078 \pm 0.003$ & -2.0\% $\pm$ 4.5\% \\
159098316 & \cite{Gandolfi2013} & $0.096 \pm 0.004$ & -3.6\% $\pm$ 3.8\% \\
159725995 & \cite{Morton2016} & $0.13 \pm 0.01$ & 14.8\% $\pm$ 9.1\% \\
164786087 & \cite{Morton2016} & $0.116 \pm 0.004$ & -8.2\% $\pm$ 4.0\% \\
164892194 & \cite{Esteves2015} & $0.092 \pm 0.003$ & 4.1\% $\pm$ 2.7\% \\
171974763 & \cite{Morton2016} & $0.14 \pm 0.02$ & 24.1\% $\pm$ 12.4\% \\
268159158 & \cite{Morton2016} & $0.09 \pm 0.01$ & 3.8\% $\pm$ 15.3\% \\
268924036 & \cite{Morton2016} & $0.128 \pm 0.008$ & 9.0\% $\pm$ 5.8\% \\
269269546 & \cite{Morton2016} & $0.13 \pm 0.01$ & 3.4\% $\pm$ 7.7\% \\
271042217 & \cite{Bonomo2015} & $0.105 \pm 0.004$ & 2.3\% $\pm$ 3.5\% \\
271354351 & \cite{Howell2010} & $0.139 \pm 0.009$ & 4.3\% $\pm$ 6.3\% \\
272366748 & \cite{Morton2016} & $0.119 \pm 0.007$ & -0.6\% $\pm$ 5.6\% \\
273874849 & \cite{Desert2011} & $0.138 \pm 0.006$ & 3.5\% $\pm$ 3.9\% \\
299220166 & \cite{Hebrard2014} & $0.123 \pm 0.004$ & 1.2\% $\pm$ 2.9\% \\
378085713 & \cite{Morton2016} & $0.12 \pm 0.04$ & 24.1\% $\pm$ 23.3\% \\
399794420 & \cite{Morton2016} & $0.119 \pm 0.006$ & -1.3\% $\pm$ 4.9\% \\
405717754 & \cite{Esteves2015} & $0.124 \pm 0.001$ & 1.9\% $\pm$ 1.1\% \\
\hline
\multicolumn{4}{l}{\textbf{Legend:}} \\
\multicolumn{4}{l}{Same as Table~\ref{tab:tess-free}, but for Kepler planets.} \\
\end{longtable}

\onecolumngrid
\section{Simplified $M$-$R$ relations model} \label{app:mr_relation}

We modeled observed exoplanet $M$-$R$ relations using a piecewise power-law function with two distinct mass regimes. The literature model \citep{Muller2024} was fitted to a subset of the PlanetS catalog, which contains mostly planets measured with Kepler \citep{Borucki2010} and ground-based telescopes. Since the literature model with movable breakpoints cannot converge on our sample with few small planets ($M_{\text{p}} < 4.37\,\text{M}_\oplus$), we simplified our model with one fixed breakpoints and removed small planets. We define the relation between planet radius ($R_{\text{p}}$) and planet mass ($M_{\text{p}}$) as
\begin{equation}
R_{\text{p}}(M_{\text{p}}) =
\begin{cases} 
c_1 M_{\text{p}}^{\alpha_1}, & 4.37\,\text{M}_\oplus <M_{\text{p}} < M_1 \\
c_2 M_{\text{p}}^{\alpha_2}, & M_{\text{p}} \geq M_1
\end{cases}
\end{equation}
where $c_1, \alpha_1, c_2, \text{and}\, \alpha_2$ define the normalization and power-law exponents in each component, and $M_1$ denote the transition masses between regimes. We set $M_1 = 127 \,\text{M}_\oplus$ according to the literature model's break point between intermediate-mass and large planets \citep{Muller2024}. 

To fit the model, we perform an orthogonal distance regression (ODR) in log-log space, transforming the data as $ \log_{10} M_{\text{p}} $ and $ \log_{10} R_{\text{p}} $. Measurement uncertainties ($ \sigma_{M_{\text{p}}} $, $ \sigma_{R_{\text{p}}} $) are propagated to logarithmic space using  

\begin{equation}
\sigma_{\log M_{\text{p}}} = \frac{\sigma_{M_{\text{p}}}}{M_{\text{p}} \ln 10}, \quad \sigma_{\log R_{\text{p}}} = \frac{\sigma_{R_{\text{p}}}}{R_{\text{p}} \ln 10}.
\end{equation}

We employ the ODR implementation from the SciPy \citep{Virtanen2020} package, fitting the model to the data with an initial parameter guess of [$c_1$, $\alpha_1$, $c_2$, $\alpha_2$] = [0.56, 0.67, 18.6, -0.06]. The optimization minimizes orthogonal residuals in log-log space, providing best-fit parameters along with their formal uncertainties. The resulting best-fit parameters for the model with literature radii are  
\begin{equation}
R_{\text{p}}(M_{\text{p}}) =
\begin{cases} 
(0.40 \pm 0.11)\, M_{\text{p}}^{(0.71 \pm 0.07)}, & 4.37\,\text{M}_\oplus < M_{\text{p}} < 127\,\text{M}_\oplus \\
(12.89 \pm 1.21)\, M_{\text{p}}^{(0.00 \pm 0.02)}, & M_{\text{p}} \geq 127\,\text{M}_\oplus
\end{cases}
\end{equation}
For the model with TGLC-fitted radii, we have
\begin{equation}
R_{\text{p}}(M_{\text{p}}) =
\begin{cases} 
(0.50 \pm 0.13)\, M_{\text{p}}^{(0.69 \pm 0.07)}, & 4.37\,\text{M}_\oplus < M_{\text{p}} < 127\,\text{M}_\oplus \\
(15.63 \pm 1.49)\, M_{\text{p}}^{(-0.02 \pm 0.02)}, & M_{\text{p}} \geq 127\,\text{M}_\oplus
\end{cases}
\end{equation}
The latter result is closer to the literature model, especially for the intermediate-mass planets. 

\bibliography{main}{}
\bibliographystyle{aasjournalv7}
\end{document}